\documentstyle[prl,aps,twocolumn,epsf]{revtex}
\draft

\begin{document}

\title{Ergodic Properties of Local Spectral Density for a
Conservative System of Coupled Quantum States}

\author{V. S. Starovoitov}

\address {Institute of physics,
NBAS, 220072, Scarina ave. 70, Minsk,Belarus}

\date{\today}

\maketitle
\begin{abstract}
The shape and the inverse participation ratio (IPR) of local spectral
density (LSD) are studied for a generic isolated system of coupled quantum
states, the Hamiltonian of which is represented by a band random matrix with
the disordered leading diagonal. We find for the matrices with arbitrary
small band that the lack of ergodicity for LSD can be associated with an
exponential increase in IPR with the ratio $v/\Delta _c$ ($v$ - the root of
mean square for off-diagonal matrix elements, $\Delta _c$ - the energy
spacing between directly coupled basis states). Criterions specifying
transition to localization and ergodicity for LSD are considered.
\end{abstract}
\pacs{PACS numbers: 05.45.Mt, 03.65.Yz, 34.30.+h, 71.55.Jv, 72.10.Bg}

\narrowtext

\section{Introduction}

The statistical properties of quantum system attract considerable attention
in the broad field of modern physics ranging from nuclear, atomic and
molecular physics to condensed matter and quantum computing. Of a special
interest is the properties specifying the quantum chaos border for a system
of many-body quantum states. Extensive investigations of many-body
interacting systems such as nuclei, many-electron atoms, quantum dots,
quantum spin glasses and quantum computer models \cite
{crossover70,Aberg90,crossover94,crossover96,crossover97,QCBorderShepel97,Weinmann97,crossover98,QuantComp}
have shown that the border is associated with a crossover of level spacing
statistics from the Poisson distribution to Wigner-Dyson one and with a
transition to ergodic distribution for the eigenstate wavefunction over the
large number of basis levels as in Random Matrix Theory (RMT) \cite
{RMT1,RMT2}. In a sense the interaction leads to dynamic thermalization
without coupling to an external thermal bath. According to \cite
{Aberg90,QCBorderShepel97,Aberg92} this crossover takes place when the
coupling matrix elements of the Hamiltonian become comparable to the energy
spacing between directly coupled states.

We address this problem to a generic system of coupled quantum states, the
Hamiltonian matrix ${\bf H}$ of that includes the leading diagonal with
disordered random values and the random off-diagonal elements inside the
band of size $b$. This band random matrix with the disordered diagonal
(BRMDD) is a reasonable model describing the systems with strong
imperfection in the basis state energies. Models based on BRMDD were applied
to study the electron transport problem \cite
{ElectrTransport82,ElectrTransport95} and the problem of interacting
particles in a random potential \cite{BRMPB94,BRMPB95,Weinmann96}. The
results obtained for these models are of an obvious interest also in
analyzing such few-freedoms physical objects as the vibrational
quasicontinuum of polyatomic molecule \cite{MolVibr} and a quantum computer 
\cite{QuantComp}.

The statistical properties of BRMDD-based systems were studied in some
details \cite{BRMPB94,BRMPB95,BRMPB91,BRMPB94a,x0,x1,BRMPBtheor95,x2}. The
shape, the localization length and the inverse participation ratio (IPR) for
eigenfunction have been investigated with the help of numerical simulations 
\cite{BRMPB94,BRMPB95,x1} and the supersymmetry approach \cite
{x1,BRMPBtheor95}. These investigations have exhibited the Lorentzian shape
for local spectral density (LSD) in circumstances when a non-perturbative
localization regime is realized. The state-state interaction strength, at
which the eigenstates are extended over the whole matrix size $N$ and the
eigenenergy level spacing statistics has the Wigner-Dyson form, have been
revealed for BRMDD with sufficiently large band (when $2b+1\gg \sqrt{N}$) 
\cite{BRMPB95,x0,x1,BRMPBtheor95}. Unfortunately the ergodic properties of
LSD have not been investigated in details. An open question is also the
quantum chaos border for BRMDD with arbitrary small band.

In the paper we study localization and ergodicity properties of LSD to be
obtained from BRMDD with arbitrary small band. The LSD was introduced in
1955 by Wigner \cite{Wigner55} and successfully employed in RMT to describe
statistically the localization effects for complex quantum systems \cite
{RMT1,RMT2} (including the systems, represented by band random matrices with
the reordered leading diagonal \cite{Casati93,Cohen00}). This quantity is
the Fourier transform for the correlation function $C(t)=\left\langle
0\right| \exp (-i{\bf H}t/\hbar )\left| 0\right\rangle $ and specifies
spreading of the energy, initially concentrated in a specific unperturbed
basis state $\left| 0\right\rangle $, between the eigenstates due to
state-state interaction. Generally the LSD may be characterized by the width 
$\Gamma $ measuring the energy scale, in that the individual state $\left|
0\right\rangle $ is localized. The number of eigenstates populating this
scale is given in terms of the product

\begin{equation}
\label{ErgLength}\xi _e=\rho _E\Gamma 
\end{equation}
designated here as {\it an ergodic localization length} of LSD ($\rho _E$ is
the eigenstate density). Hence the quantity $\xi _e$ specifies the greatest
possible number of eigenstates, where the basis state $\left| 0\right\rangle 
$ can be effectively admixed. The ergodic properties for the system can be
identified with the structure of LSD. The non-ergodic LSD is a strongly
fluctuating spiked function and the IPR $\xi _{IPR}$, which gives the actual
number of eigenstates involving the state $\left| 0\right\rangle $, is low
in comparison with $\xi _e$. In the ergodicity case the LSD is monotonic and
the number $\xi _{IPR}$ approaches the value of $\xi _e$.

The main attention of our study is paid to the perturbative and localized
regimes for LSD. Therefore we restrict the study to a moderate strength of
state-state interaction. This restriction implies that the width $\Gamma $
is essentially small as compared to the energy scale bounding the location
of the eigenstate levels ($\xi _{IPR},\xi _e<<N$). The finite-size effects
are ignored. Hence the quantities $\xi _e$ and $\xi _{IPR}$ can be
considered as functions of the interaction strength and the relative width
of band.

We investigate the LSD shape and determine the ergodic localization length.
Then we analyze the behavior of IPR for ergodic and non-ergodic LSD. From
this analysis we obtain criterions specifying the transition to localized
and ergodic regimes for LSD.

\section{Model description}

We consider a BRMDD, that represents the Hamiltonian matrix ${\bf H}$ in the
basis of unperturbed states $\left| k\right\rangle $ ($k=-K,\ldots ,K$) for
isolated system of $N=2K+1$ quantum states.The BRMDD is a real symmetric
matrix with statistically independent random elements 
\begin{equation}
\label{Hamilt}H_{mk}=E_k^{(0)}\delta _{mk}+V_{mk} 
\end{equation}
where off-diagonal elements $V_{mk}=V_{km}$ specify the state-state
interaction. The values of $V_{mk}$ are distributed uniformly in the
interval $[-V,V]$ with $\langle V_{mk}\rangle =0$ and $v^2=\langle
V_{mk}^2\rangle =V^2/3$ if $\left| m-k\right| \leq b$ or are zero otherwise.
The diagonal elements $E_k^{(0)}$ corresponding to energy levels for the
states $\left| k\right\rangle $ are uniformly distributed according to the
Poisson statistics with the mean spacing $\Delta $ between adjacent energy
levels: $-K\Delta \leq E_k^{(0)}\leq K\Delta $. The energy level $E_0^{(0)}$
for a probing state $\left| 0\right\rangle $ is located in the midpoint of
interval $[-K\Delta ,K\Delta ]$: $\langle E_k^{(0)}\rangle \approx
E_0^{(0)}=0$.

The perturbation of an individual basis state depends on the relative level
position for unperturbed states, those are coupled directly with the state
by interaction. Therefore the strength of state-state interaction for BRMDD
can be specified in terms of the ratio $v/\Delta _c$, where 
\begin{equation}
\label{DeltaC}\Delta _c=\frac \Delta \beta 
\end{equation}
is the energy spacing between directly coupled unperturbed states. Here the
relative band width 
\begin{equation}
\label{beta}\beta =\frac bK 
\end{equation}
gives the relative number of states coupled directly with the individual
state.

Depending on the relation between $v/\Delta _c$ and $b$, three important
regimes can be distinguished \cite{BRMPB94,BRMPB95,x1,BRMPBtheor95}. When
the coupling matrix elements are weak and $v/\Delta _c\ll 1$ the
perturbative regime is realized. At the regime the Hamiltonian (\ref{Hamilt}%
) can be treated within the framework of perturbation theory. The condition $%
1\ll v/\Delta _c\ll \sqrt{b}$ features the regime, at that the eigenstates
are localized. The strong interaction at $v/\Delta _c\gg \sqrt{b}$ is
responsible for the delocalized regime for eigenfunction. In the regime the
eigenstates are extended over the whole matrix size $N$.

The eigenfunctions $\left| \alpha _n\right\rangle $ and the eigenenergies $%
E_n$ are obtained from diagonalization of (\ref{Hamilt}). The quantity $%
W_{n0}=\left| \left\langle \alpha _n|0\right\rangle \right| ^2$ gives a
probability to find the probing state $\left| 0\right\rangle $ in the
eigenstate $\left| \alpha _n\right\rangle $. The numerical simulation is
performed for the wide range of matrix parameters: $2<N<3200$, $1<b<1600$
and $10^{-3}<v/\Delta <10^2$. The number of disorder realizations lies in
the range from 100 to 1000.

\section{Shape of LSD}

\begin{figure} [h!]
\epsfxsize= 8 cm
\centerline{\epsfbox{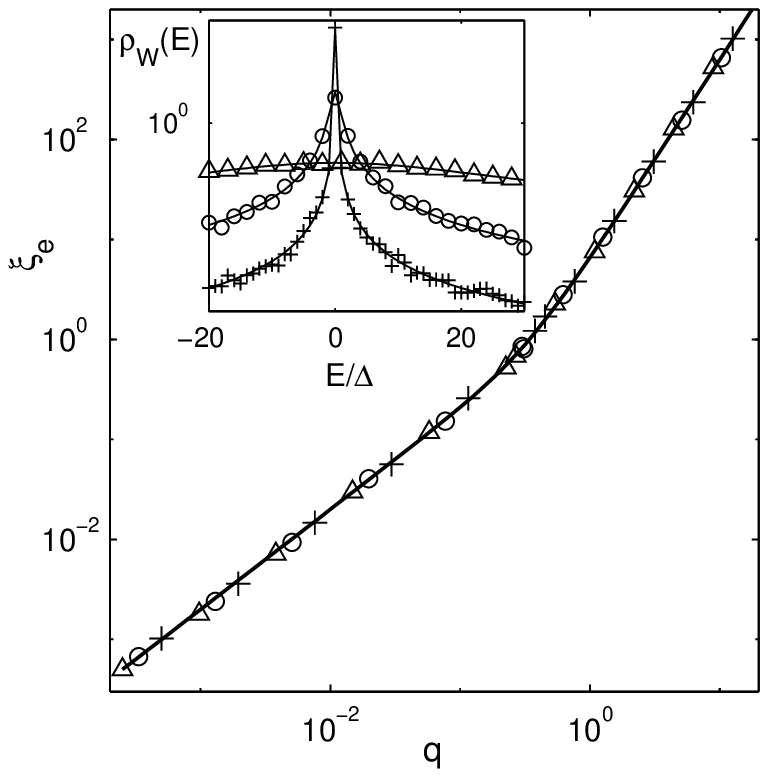}}
\vspace{3 mm}
\caption{
Dependence of the ergodic localization length $\xi _e$ on $q$
for $\beta =$1 (+), 0.5 ($\bigcirc $) and 0.04 ($\bigtriangleup $).
The solid line gives the approximation (9).
The inset shows a semilog plot for the LSD
$\rho _W(E)$ for $\beta =1$ at $v/\Delta _c =0.089$
($+$, 1000 disorder realizations), $v/\Delta _c =0.45$
($\bigcirc $,100 disorder realizations) and $v/\Delta _c =2.82$
($\bigtriangleup $, 100 disorder realizations). The solid lines give the
Breit-Wigner distribution (6) with $\xi _e=$0.05, 1.3
and 50.
}
\label{FigL(q)}
\end{figure}

The shape of LSD is studied for the eigenenergy scale, which is no larger
than $5\Gamma $. For this scale the interaction-induced variations in level
density are negligible and the eigenenergies are homogeneously distributed
with the density $\rho _E\approx 1/\Delta $. Then the LSD can be defined as 
\begin{equation}
\label{LSD}\rho _W\left( E\right) =\rho _E\frac{\left\langle
\sum_nW_{n0}\delta (E-E_n)\right\rangle }{\left\langle \sum_n\delta
(E-E_n)\right\rangle }, 
\end{equation}
where $\langle \ldots \rangle $ means the averaging over disorder (that is,
over many random matrices). The numerical investigation confirms that the
LSD is described by the well-known Breit-Wigner distribution of the
Lorentzian shape 
\begin{equation}
\label{BWdistr}\rho _{BW}(E)=\frac \Gamma {2\pi }\frac 1{E^2+\Gamma ^2/4} 
\end{equation}
(see the inset in Fig.\ref{FigL(q)}) with the width $\Gamma $. This shape is
observed for any $v$, $\Delta _c$, $b$ and $N$ satisfying $\xi _e\ll N$
(including the parameter region where $\Gamma \ll \Delta $). We find the
ergodic localization length $\xi _e$ from (\ref{ErgLength}) where the width $%
\Gamma $ is estimated by fitting of the Lorentzian (\ref{BWdistr}) to an
averaged LSD. Our analysis of numerical data obtained for different values
of $v$, $\Delta _c$ and $\beta $ shows that the quantity $\xi _e$ can be
considered as a function of the single parameter 
\begin{equation}
\label{q}q=\frac v{\Delta _c\sqrt{\beta }}, 
\end{equation}
which defines essentially the localization of LSD in the eigenenergy scale.
When the coupling of the basis states by perturbation is sufficiently strong
and $q>>1$ the quantity $\xi _e$ satisfies the Fermi golden rule: $\xi
_e\Delta =\Gamma \approx 2\pi q^2\Delta =2\pi v^2/\Delta _c\gg \Delta $. At
low values of $q$ ($q<<1$) the length $\xi _e$ can be described by a linear
function of $q$: $\xi _e\approx 2q$. The width $\Gamma $ is then much
smaller as compared to the spacing $\Delta $. Fig.\ref{FigL(q)} demonstrates
the behavior of $\xi _e$ in the wide range of $q$ at three fixed magnitudes
of $\beta $.

The linear dependence of $\xi _e$ and $\Gamma $ on the strength of
state-state coupling for small $q$ can be explained in terms of a simple
perturbation approach to the eigenenergy levels $E_n$ located in the
vicinity of $E_0^{(0)}$: $\left| E_n-E_0^{(0)}\right| \lesssim v\ll \Delta $%
. In the approach the eigenstate $\left| \alpha _n\right\rangle $ is
considered as a superposition of the nearest basis states $\left|
0\right\rangle $ and $\left| n\right\rangle $. The probability $W_{n0}$ to
find the basis state $\left| 0\right\rangle $ in the eigenstate $\left|
\alpha _n\right\rangle $ can be then estimated from a two-state (the states $%
\left| 0\right\rangle $ and $\left| n\right\rangle $) approximation.
According to this approximation the dependence of $W_{n0}$ on $E_n$ for an
individual disorder realization is Lorentzian-like 
\begin{equation}
\label{Wn0}W_{n0}\approx \frac{V_{n0}^2}{V_{n0}^2+\left(
E_n-E_0^{(0)}\right) ^2} 
\end{equation}
with the width of $2\left| V_{n0}\right| $. If the states $\left|
0\right\rangle $ and $\left| n\right\rangle $ are not coupled by interaction
the dependence (\ref{Wn0}) is considered as a function, the width of that is
negligible in comparison to $v$. The probability that the states $\left|
0\right\rangle $ and $\left| n\right\rangle $ are coupled by interaction is
equal to $\beta $. As a result the shape of disorder-averaged LSD is
described by the Lorentzian-like contour with the width $\Gamma \approx
2\beta v$. It seems that this perturbation approach can be applied also for
the eigenstates with $\left| E_n-E_0^{(0)}\right| \gg \Delta $, when the
probability $W_{n0}$ is essentially small as compared to the contributions
of others basis states. In this limit Eq. (\ref{Wn0}) gives $W_{n0}\approx
V_{n0}^2/(E_n^{(0)}-E_0^{(0)})^2$, as predicted in the framework of a
standard perturbation approach for non-degenerate states (see, for instance,
in \cite{QuantMech}).

In order to estimate the localization properties of LSD at intermediate
values of $q$ we approximate the behavior of ergodic length for the wide
range of $q$ ($10^{-3}<q<15$). The analysis of numerical data shows that the
length $\xi _e$ can be well approximated by 
\begin{equation}
\label{L(q)}\xi _e\approx L_1q\sqrt{1+(L_2q)^2} 
\end{equation}
with the fitted coefficients $L_1=2.01\approx 2$ and $L_2=3.16\approx \pi $.
According to (\ref{L(q)}) an accurate boundary between the linear and
quadratic dependencies of $\xi _e$ on $q$ is obtained at $q\approx 1/L_2$
where $\xi _e\approx 0.9$. Notice that this boundary separates the ranges
where the value of $\xi _e$ is low or high as compared to $1$.

\section{Participation ratio for LSD}

The IPR $\xi _{IPR}=(\langle \sum_n\mid W_{n0}^{(s)}\mid ^2\rangle _s)^{-1}$
is associated with the parameters $q$ and $\beta $, which specify
correspondingly the ergodic localization length $\xi _e$ for LSD and the
band width for BRMDD. This association depends on whether or not the LSD is
ergodic. In the ergodic case the probing state is monotonically spread over
eigenstates in the energy scale $\Gamma $. Then the IPR is a function of the
only parameter $q$ and approaches the ergodic length $\xi _e$. The ergodic
LSD can be realized from BRMDD with large band. Our analysis of calculated
data for the matrices with $\beta \sim 1$ shows that the best approximation
for ergodic IPR is 
\begin{equation}
\label{IPRerg}\xi _{IPR}\approx 1+D_1q\sqrt{1+(D_2q)^2}\approx 1+\xi _e. 
\end{equation}
The fitted coefficients $D_1=3.16$ and $D_2=1.94$ are obtained for full
Hamiltonian matrices (\ref{Hamilt}) with $\beta =1$ where all $2K$ basis
states are directly linked with the state $\left| 0\right\rangle $. Notice
that the approximation (\ref{IPRerg}) gives the maximum allowable magnitude
of IPR for BRMDD at fixed $q$.

\begin{figure} [h!]
\epsfxsize= 8 cm
\centerline{\epsfbox{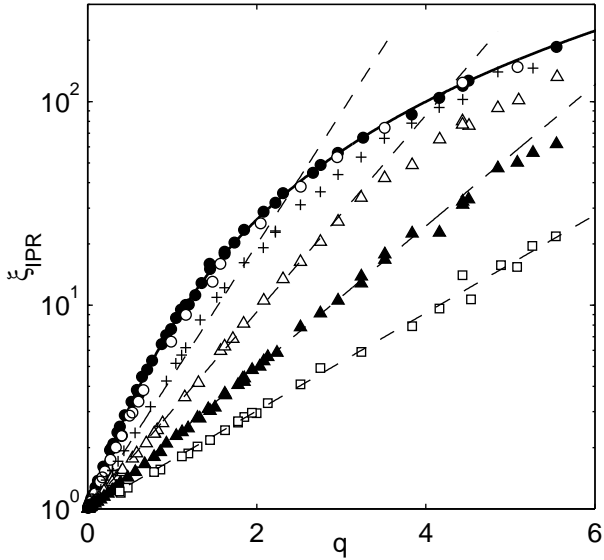}}
\vspace{3 mm}
\caption{
Dependence of the IPR $\xi _{IPR}$ on
$q$ for $\beta =$1 (full circle), 1/2 ($\bigcirc $), 1/4 ($+$), 1/8 ($\bigtriangleup $),
1/16 (full triangle) and 1/32 ($\Box $).
The dashed lines give the exponential (11)
for $\beta =1/4$ ($C=2.99$), $1/8$ ($C=3.15$), $1/16$ ($C=3.19$)
and $1/32$ ($C=3.13$).
The solid line shows the ergodic approximation (6) of IPR.}
\label{FigIPRonQ}
\end{figure}

An essentially non-ergodic LSD can be obtained from BRMDD with $\beta \ll 1$%
. Then the IPR is significantly low in comparison to $\xi _e$ and should be
considered as a function depending both on the parameter $q$ and on $\beta $%
. The analysis of numerical data for matrices with low magnitudes of $q$ and 
$\beta $ shows that the lack of ergodicity for LSD can be clearly identified
with an exponential increase in IPR with the quantity $q\sqrt{\beta }%
=v/\Delta _c$%
\begin{equation}
\label{IPRnonerg}\xi _{IPR}\approx \exp \left( Cq\sqrt{\beta }\right) . 
\end{equation}
At fixed $\beta $ the IPR is shown in Fig.\ref{FigIPRonQ} to increase
exponentially with the rate $C\sqrt{\beta }$ ($C\approx 3.0\div 3.2$) as the
parameter $q$ rises if $\xi _{IPR}\ll \xi _e$. The increase is observed up
to the point where $\xi _{IPR}\sim \xi _e$. Then the IPR approaches
asymptotically the length $\xi _e$ from below with increasing $q$. At high
magnitudes of $q$ and $\beta $, when the exponential (\ref{IPRnonerg}) gives
higher values of $\xi _{IPR}$ in comparison to $\xi _e$, the LSD is ergodic
and the IPR can be determined from (\ref{IPRerg}).

\begin{figure} [h!]
\epsfxsize= 7.5 cm
\centerline{\epsfbox{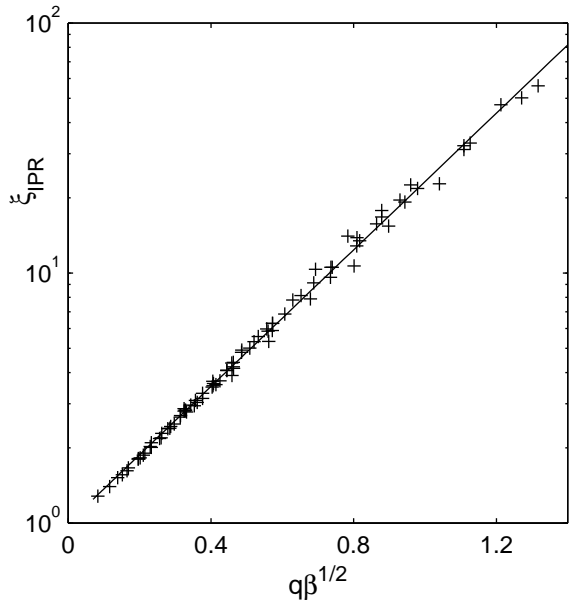}}
\vspace{3 mm}
\caption{
Dependence of the IPR $\xi_{IPR}$ on the product
$q\sqrt{\beta }$
for the broad range of $q$ ($0.01<q<10$)
and $\beta$ ($\beta >0.001$) satisfying the requirement
$\xi _{IPR}<\xi _{e}/2.7$ (+).
Solid line shows the approximating exponential (11) with
$C=3.15$.}
\label{FigIPRonQB}
\end{figure}

For the broad range of $q$ and $\beta $ satisfying to $\xi _{IPR}<\xi _e/2.7$
the mean square fitting gives the coefficient $C=3.15\pm .01$. Fig.\ref
{FigIPRonQB} demonstrates a good accordance between the calculated and
fitted values of IPR for non-ergodic LSD. Unfortunately we can give no heuristic
explanation of the exponential rise in IPR with $q\sqrt{\beta }$.
Notice only that this rise is observed under conditions when the eigenstates
are essentially localized and $l_{sb}/N\approx q^2\beta ^2\ll 1$ ($l_{sb}$
is the localization length for eigenstate).

\section{Criterions of localization and ergodicity of LSD}

We associate the localized regime for LSD with a high magnitude of IPR. For
BRMDD with arbitrary small band the requirement $\xi _{IPR}\gg 1$ is seen
from (\ref{IPRerg}) and (\ref{IPRnonerg}) to result in 
\begin{equation}
\label{QCBorder}v/\Delta _c=q\sqrt{\beta }>q_d\sqrt{\beta _d}=1/C_0\sim
1/\pi , 
\end{equation}
no matter what the ergodic properties of LSD and the magnitude of $\beta $.
The parameters $q_d$ and $\beta _d$ defines then a localization border on
the plane ($q,\beta $), that is a boundary between regions where the
perturbative ($\xi _{IPR}\sim 1$) or localized ($\xi _{IPR}\gg 1$) regimes
are realized. Notice that as for LSD, the criterion (\ref{QCBorder})
specifies a transition to the localized regime for eigenstate \cite
{BRMPB94,BRMPB95,x1,BRMPBtheor95}.

The attainment of ergodic LSD for BRMDD is defined by the relative width of
band. For the matrices with $\beta \sim 1$ the localized LSD is nearly
ergodic and the ergodicity criterion can be represented by the condition (%
\ref{QCBorder}). This condition is in good agreement with a quantum chaos
border for many-body systems and gives the crossover of level spacing
statistics from Poisson to Wigner-Dyson distribution for a nucleus model ($%
C_0=3.0$ \cite{Aberg92}) and for a generic model of quantum computer ($%
C_0=2.5$~\cite{QCBorderShepel97}). In (\ref{QCBorder}) the factor $C_0$ is
assumed to be equal to $C\approx D_1\approx \pi $.

For BRMDD with $\beta \ll 1$ we should take into account the exponential
growth (\ref{IPRnonerg}) of IPR with $v/\Delta _c$. The criterion of LSD
ergodicity can be determined then from the requirement $\exp (Cq\sqrt{\beta }%
)>\xi _e$, that results in: 
\begin{equation}
\label{QEBorder}v/\Delta _c=q\sqrt{\beta }>q_e\sqrt{\beta _e}=\frac{\ln
\left( 2\pi q_e^2\right) }C \approx \frac{\ln\left( 2\pi q_e^2\right) }\pi ,
\end{equation}
where the values of $q_e$ and $\beta _e$ specify an ergodicity border on the
plane ($q,\beta $). Notice that the condition (\ref{QEBorder}) is a more
severe restriction imposed on the parameters $q$ and $\beta $ than the
criterion (\ref{QCBorder}).

\begin{figure} [h!]
\epsfxsize= 8 cm
\centerline{\epsfbox{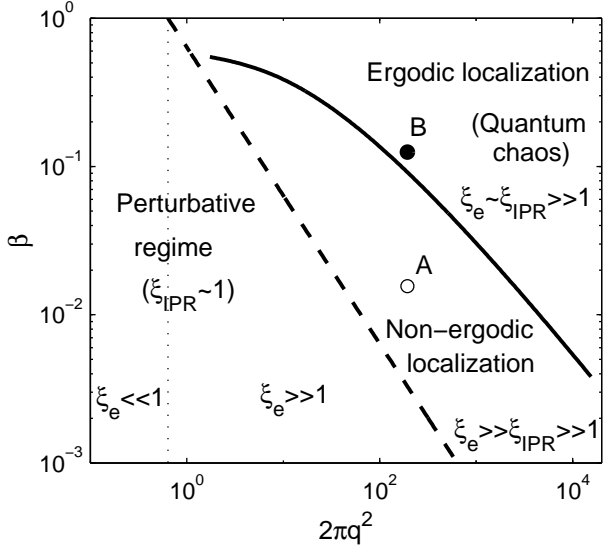}}
\vspace{3 mm}
\caption{
Localization regimes depending on the parameters $q$ and $\beta $.
The fat dashed line shows the localization border (12).
The fat solid curve gives the LSD ergodicity border (13).
The dotted line indicates a boundary between the regions
with $\xi _e<1$ and $\xi _e>1$.
The points A ($q=5.55, \beta =0.016$) and B ($q=5.55, \beta =0.125$)
give the parameter sets used to obtain the level spacing statistics
represented in Fig.5.}
\label{FigRegimes}
\end{figure}

Our study reveals some distinctive features in behavior of LSD for BRMDD
with arbitrary small band. These features are outlined on the plane ($%
q,\beta $) in Fig.\ref{FigRegimes} to depend on the relation between $%
v/\Delta _c$ and $\beta $. As for eigenstate, the perturbative regime for
LSD can be associated with the requirement $v/\Delta _c\ll 1/3$. In contrast
to eigenstate, the localization properties of LSD is essentially defined by
the magnitude of $q$. In the limit $q\ll 1/3$ (that is, $v/\Delta _c\ll 
\sqrt{\beta }/3$) the ergodic localization length is a linear function of $q$
and $\xi _e\approx 2v/\Delta _c\sqrt{\beta }$. Due to weak state-state
interaction the LSD is a delta-like function confined in an eigenstate: $\xi
_e\ll \xi _{IPR}\sim 1$. At higher values of $q$ ($1\ll 3q\ll 1/\sqrt{\beta }
$ or $\sqrt{\beta }\ll 3v/\Delta _c\ll 1$) the quantity $\xi _e$ satisfies
the Fermi golden rule: $\xi _e\approx 2\pi v^2/\Delta _c^2\beta $. Despite
the wide spreading of basis states over the eigenenergy scale, the actual
number of eigenstates associated with an individual basis state for this
parameter region remains small: $\xi _e\gg \xi _{IPR}\sim 1$. 

\begin{figure} [h!]
\epsfxsize= 7.5 cm
\centerline{\epsfbox{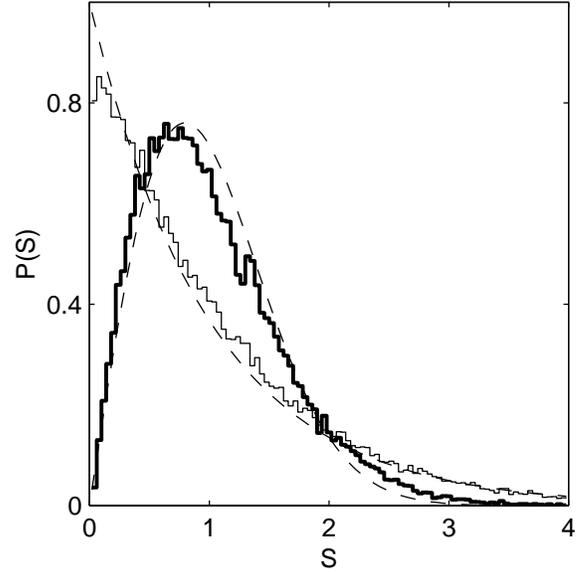}}
\vspace{3 mm}
\caption{
Level spacing statistics $P(S)$ at $\xi _e=193.5$
for non-ergodic
($\beta =0.016$, $\xi _{IPR}=10.3$, solid line stairs)
and ergodic
($\beta =0.125$, $\xi _{IPR}=150.5$, fat line stairs)
LSD localization regimes,
represented correspondingly by the points A and B in Fig.4.
The dashed curves show the Poisson and Wigner-Dyson distributions.}
\label{FigSpacing}
\end{figure}

The criterion (\ref{QCBorder}) specifies a transition from the perturbative
to localized regime both for eigenstate and for LSD. Depending on the
ergodicity properties of localized LSD, one can distinguish two important
parameter ranges. An area, located between the borders (\ref{QCBorder}) and (%
\ref{QEBorder}) on the plane ($q,\beta $), corresponds to the parameter
region $1\ll 3v/\Delta _c\ll \ln (\xi _e)$ where $\xi _e\approx 2\pi
v^2/\Delta _c^2\beta >>\xi _{IPR}\approx \exp (Cv/\Delta _c)>>1$. For this
region the neighboring eigenstates are weakly coupled by interaction since
the density $\xi _{IPR}/\Gamma $ of eigenstates, those involve the probing
state, is negligible in comparison to the total state density $\rho _E=\xi
_e/\Gamma $. It means that the level spacing statistics $P(S)$ should
exhibit the Poisson-like distribution as it shown in Fig.\ref{FigSpacing}.
We identify this region with a non-ergodic localized regime for LSD. The
regime of ergodic localization for LSD at $v/\Delta _c\gg \ln (\xi _e)/3$ is
represented by a domain above the ergodicity border (\ref{QEBorder}) where 
$\xi _e\approx 2\pi v^2/\Delta _c^2\beta \sim \xi _{IPR}>>1$.
For this domain
the level spacing statistics is close to the Wigner-Dyson distribution (see
Fig.\ref{FigSpacing}) to be associated usually with quantum chaos. In a
sense the condition (\ref{QEBorder}) can be considered as a quantum chaos
criterion.

\section{Conclusion}

We have analyzed the properties of LSD for a generic conservative system of
coupled states, the Hamiltonian of which is represented in terms of a BRMDD.
As for eigenstate, the transition from the perturbative to localized regime
for LSD can be associated with the same requirement to be imposed on the
strength of state-state interaction. Nevertheless, our study has revealed
some distinctive features in the properties of LSD. We have determined
localization and ergodicity borders for LSD obtained from BRMDD with
arbitrary small band. For the matrices with large band ($\beta \sim 1$) the
borders coincide with each other and are in good agreement with the quantum
chaos border for some generic many-body systems. For the matrices with small
band ($\beta \ll 1$) the criterion of LSD ergodicity is a more severe
requirement imposed on the band size and the interaction strength than the
localization criterion. In the parameter range between the borders the IPR
is found to increase exponentially with the ratio $v/\Delta _c$, which
specifies the strength of state-state interaction. We suppose that such an
effect can be observed for any isolated system of coupled states with strong
imperfection in non-perturbed energies.

Author thanks D. Shepelyansky and V. Churakov for useful discussions and
remarks.

\end{document}